\documentclass[a4paper,twocolumn,prb,amsmath,amssymb,floatfix,preprintnumbers]{revtex4}

\newcommand{\veps}{\varepsilon}

\usepackage{graphicx}
\usepackage{dcolumn}
\usepackage{bm}

\begin{document}


\title{
Large-scale electronic structure calculation and its application
}
\author{Takeo Hoshi
\footnote{http://fujimac.t.u-tokyo.ac.jp/hoshi/}}
\affiliation{
Department of Applied Physics, 
University of Tokyo, Bunkyo-ku,
Tokyo, Japan 
}

\begin{abstract}
Several methodologies are developed
for large-scale atomistic simulations with 
fully quantum mechanical description of electron systems. 
The important methodological concepts 
are (i) generalized Wannier state, 
(ii) Krylov subspace
and (iii) 
hybrid scheme within quantum mechanics.
Test calculations are done 
with upto 10$^6$ atoms using a standard workstation. 
As a practical nanoscale calculation,
the dynamical fracture of 
nanocrystalline silicon was simulated. 

\end{abstract}

\maketitle


\section{Introduction}

Nanoscale materials
are directly governed by 
quantum mechanical freedoms 
of electron systems.
Nowadays,
electron systems of realistic materials 
are treated by 
the {\it ab initio} electronic structure calculations
that are based on 
the density functional theory (DFT) 
\cite{HOHENBERG-KOHN,KOHN-SHAM},
the first-principle molecular dynamics
\cite{CP},
and related theories developed for decades. 
A typical system size of 
present {\it ab initio} calculations is, 
however, 
on the order of 10$^2$ atoms and
new practical theories are required
for nanoscale calculations.
This article is devoted to 
the methods in large-scale electronic structure calculations
and their application to nanoscale materials 
\cite{Hoshi00a,Hoshi01,Hoshi02b, 
GESHI,TAKAYAMA,Hoshi03a}.

In general, 
a quantum mechanical calculation of an electron system
is reduced to an eigen value equation;
\begin{eqnarray}
 \hat{H}   | \phi_k^{(\rm eig)} \rangle =
 \varepsilon_k^{\rm (eig)}  | \phi_k^{(\rm eig)} \rangle
 \label{EIGEN-EQ}
\end{eqnarray}
with an effective one-body Hamiltonian $\hat{H}$.
Here the eigen energies and eigen states are denoted as
$\{ \varepsilon_k^{\rm (eig)} \}$ and
$\{  \phi_k^{(\rm eig)}  \}$, respectively. 
A physical quantity $\langle \hat{X} \rangle $ is given as
\begin{eqnarray}
\langle \hat{X} \rangle = \sum_k^{\rm occ.} 
\langle \phi_k^{(\rm eig)} | \hat{X}  | \phi_k^{(\rm eig)} \rangle 
= {\rm Tr}[\hat{\rho} \hat{X}]
 \label{TRACE-EQ}
\end{eqnarray}
with occupied eigen states $\{\phi_k^{\rm (eig)}\}$ or the one-body density matrix $\hat{\rho}$
\begin{eqnarray}
\hat{\rho} \equiv \sum_k^{\rm occ.}  |   \phi_k^{\rm (eig)} \rangle  \langle  \phi_k^{\rm (eig)} |.
 \label{DM-DEF}
\end{eqnarray}
The calculation of eigen states, 
Eq.~(\ref{EIGEN-EQ}),
is usually reduced to  a matrix diagonalization procedure and
gives a severe computational cost. 
Therefore, 
the essential methodology for large-scale calculations 
is how to obtain the  density matrix $\hat{\rho}$ 
{\it without calculating eigen states}. 
This article focuses the methods for structural properties 
including molecular dynamics simulations.
For the above purpose,
the most import physical quantity is the total energy 
\cite{NOTE}.

\begin{figure}[thb]
\begin{center}
  \includegraphics[width=6cm]{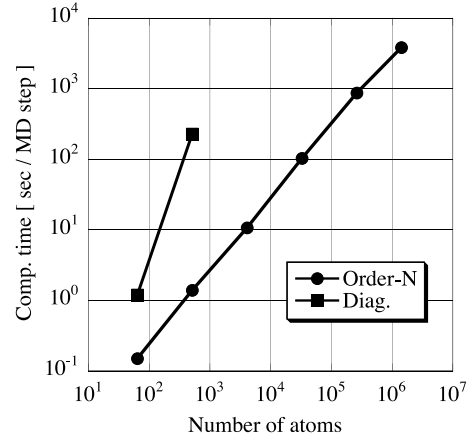}
\end{center}
\caption{
The computational time of bulk silicon as the function 
of the number of atoms ($N$),
up to 1,423,909 atoms \cite{Hoshi02b};
The CPU time is measured 
for one time step in the molecular dynamics (MD) simulation. 
A tight-binding Hamiltonian is solved using 
the exact diagonalization method and 
an \lq order-$N$' method
with the perturbative Wannier state
(See Section \ref{SEC-WANI-ON}).
We use a standard 
work station with one Pentium 4$^{\rm TM}$ processor 
and 2 GB of RAM.
}
\label{FIG-CPU-ORDER-N}
\end{figure}%

The trace in Eq.~(\ref{TRACE-EQ}) is expressed as
\begin{eqnarray}
 {\rm Tr}[\hat{\rho} \hat{X}]
=  \int \! d \bm{r} \int \! d \bm{r}' 
  \rho(\bm{r},\bm{r}') X(\bm{r}',\bm{r})
\end{eqnarray}
with real space coordinates $\bm{r}$,$\bm{r}'$.
The {\it off-diagonal} components 
of the density matrix,
$\rho(\bm{r},\bm{r}'),  \bm{r} \ne \bm{r}'$,
are essential quantum mechanical freedoms,
while the {\it diagonal} components,
the charge density at the point $\bm{r}$ 
$(\rho(\bm{r},\bm{r})\equiv n(\bm{r}))$,
appear also in classical mechanics. 
As an important fact for practical large-scale calculations,
the off-diagonal long range component of the density matrix
dose not contribute explicitly to the value of $\langle \hat{X} \rangle $,
if the operator $\hat{X}$ is a short range one.
We can found 
a general principle within DFT,
called \lq nearsightedness principle'\cite{KOHN96},
which is directly related to the above fact.

For practical algorithms,
there are many proposals. 
See reviews or comparison papers
\cite{ORDEJON-REV, Goedecker99a,GALLI-REV,JAYANTHI-REV,BOWLER97A}. 
In this article, we pick out two methods;
(i) method with generalized Wannier state and
(ii) Krylov subspace method.
Figure \ref{FIG-CPU-ORDER-N} demonstrates 
the computational cost with diagonalization and 
our calculation\cite{Hoshi02b}. 
Here one can find that 
the diagonalization results in
a compuational cost proportional to $N^3$ 
with the system size $(N)$,
as is usual in matrix diagonalization procedure 
($\propto N^3$).
Our calculation, on the other hand,
shows an \lq order-$N$' property,
with upto 10$^6$ atoms, 
in the sense that 
the computational cost is proportional to the system size 
($\propto N$).

\section{Generalized Wannier state \label{SEC-WANI-ON}}

The generalized Wannier state
is a generalization of 
the (conventional) Wannier states 
\cite{WANNIER-ORG,KOHN-WANI59,NOTE-AM}
(See Appendix \ref{APPEND}).
Its pioneering works 
were done by Walter Kohn
in the context of 
large-scale calculations
\cite{KOHN-WANI73,KOHN-WANI93}.
The pictures of 
the generalized Wannier states 
are localized \lq chemical' wave functions in condensed matter,
such as a bonding orbital or a lone-pair orbital,
with a slight spatial extension or \lq tail'.

The generalized Wannier states $\{  \phi_i^{\rm (WS)}  \}$
are defined as localized wave functions that 
satisfy the equation
\begin{eqnarray}
 H | \phi_i^{\rm (WS)} \rangle
 = \sum_{j=1}^{\rm occ} \varepsilon_{ij} | \phi_j^{\rm (WS)} \rangle
 \label{REV-SCE-UNITARY}
\end{eqnarray}
and the orthogonality
\begin{eqnarray}
 \langle \phi_i^{\rm (WS)}  | \phi_j^{\rm (WS)} \rangle = \delta_{ij}.
 \label{REV-WANI-ORTHOGO}
\end{eqnarray}
The matrix $\varepsilon_{ij}$ is introduced 
as the Lagrange multiplier 
for the constraint of Eq.~(\ref{REV-WANI-ORTHOGO})
and is given as 
\begin{eqnarray}
 \varepsilon_{ij} = 
 \langle \phi_j^{\rm (WS)} | H | \phi_i^{\rm (WS)} \rangle.
\end{eqnarray}
The solutions of Eq.~(\ref{REV-SCE-UNITARY})
is equivalent to the unitary transformation
of the eigen states $\{\phi_k^{\rm (eig)} \}$
\begin{eqnarray}
 | \phi_i^{\rm (WS)} \rangle = \sum_{k}^{\rm occ.} U_{ik}  | \phi_k^{\rm (eig)} \rangle,
  \label{WANI-REV-LOC-ST-UT}
\end{eqnarray}
where $ U_{ik}$ is a unitary matrix.
Here the suffix $i$ of the Wannier state $\phi_i^{\rm (WS)}$
denotes its localization center.

It is crucial that the generalized Wannier states
reproduce the one-body density matrix $\hat{\rho}$ in 
Eq.~(\ref{DM-DEF}), 
where the eigen states 
$\{ \phi_k^{\rm (eig)}\}$
are replaced by the Wannier states 
$\{ \phi_j^{\rm (WS)}\}$.
In results, any physical quantity can be reproduced 
in the trace form of Eq.~(\ref{TRACE-EQ}).
\cite{NOTE-WS-LMTO}.

The concept of the generalized Wannier state
is used for practical large-scale calculations
\cite{LOM-MGC,LOM-MGC2,Hoshi00a}.
We derived 
a mapped eigen value equation 
for the generalized Wannier states
\begin{eqnarray}
 H_{\rm WS}^{(i)}  | \phi _i^{\rm (WS)} \rangle 
 = \veps_{\rm WS}^{(i)}  | \phi _i^{\rm (WS)} \rangle
 \label{MFE}
\end{eqnarray}
with a mapped Hamiltonian $H_{\rm WS}^{(i)}$ 
that is dependent on the other Wannier states 
$\{\phi_j^{\rm (WS)}\}_{j \ne i} $
\cite{Hoshi00a,Hoshi01}.
Equation (\ref{MFE})
is equivalent to 
Eqs.~(\ref{REV-SCE-UNITARY}) and 
(\ref{REV-WANI-ORTHOGO}).
A Wannier state $| \phi_i^{\rm (WS)} \rangle $ is 
not an eigen state of the original Hamiltonian $H$
but an eigen state of the above mapped Hamiltonian $H_{\rm WS}^{(i)}$.
Equation (\ref{MFE}) also shows that 
the locality of a Wannier state can be mapped,
formally, to that of a virtual impurity state \cite{Hoshi00a}.

With Eq.~(\ref{MFE}),
we developed a variational method
so as to generate approximate Wannier states,
which is called
variational Wannier state method  \cite{Hoshi00a}.
As the practical procedure,
Eq.~(\ref{MFE}) is solved iteratively under 
explicit localization constraint on each Wannier state
\begin{eqnarray}
\{ \phi_i^{\rm (WS)} \}  \rightarrow \{ H^{(i)}_{\rm WS} \} \rightarrow 
\{ \phi_i^{\rm (WS)} \}  \rightarrow \{ H^{(i)}_{\rm WS} \} \rightarrow
\cdot \cdot \cdot
 \label{MFE-SEC-WANI-ON-LOOP-ITE}
\end{eqnarray}

With Eq.~(\ref{MFE}),
we also developed a perturbative method 
to generate Wannier states,
which is called perturbative Wannier state method
\cite{Hoshi00a,Hoshi01}.
This method corresponds to a non-iterative solution
of Eq.~(\ref{MFE}).
It is noteworthy that 
the perturbative Wannier state, unlike the variational one,
is localized
{\it without any explicit localization constraint},
when a short range Hamiltonian $H$ is used.

\begin{figure*}[tbh]
\begin{center}
  \includegraphics[width=14cm]{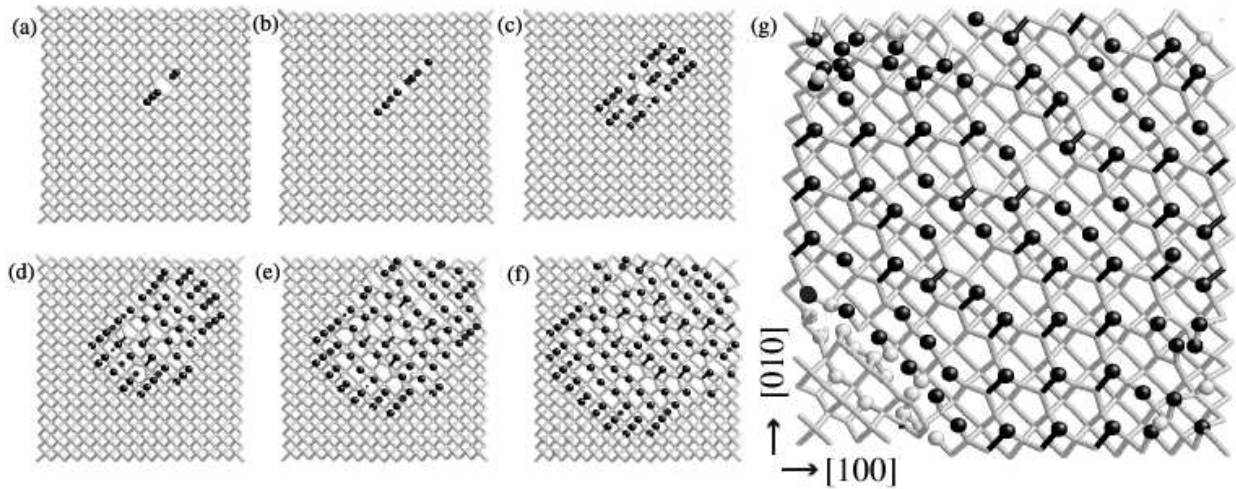}
\end{center}
\caption{
Snapshots of a fracture process in the (001) plane \cite{Hoshi02b}.
The sample contains 4501 atoms and one initial defect bond 
as the fracture {\it seed}.
The time interval between 
two successive snapshots is 0.3 ps, 
except that between (f) and (g) (approximately 1.3 ps).
A set of connected black rod and black ball
corresponds to an asymmetric dimer.
See details in the original paper \cite{Hoshi02b}.
}
\label{FIG-ANIME}
\end{figure*}

\section{Krylov subspace method}

The Krylov subspace method \cite{KRYLOV} is focused 
as another important concept for large-scale calculations. 
The Krylov subspace is a mathematical concept and 
its definition is the linear space that 
is constructed from the following vectors;
\begin{equation}
| i \rangle, \quad
H| i \rangle, \quad
H^2| i \rangle, \quad \cdot \cdot \cdot
H^{\nu-1}| i \rangle.
\label{KRY-BASIS}
\end{equation}
Here an initial vector $| i \rangle$ should be given.
In the present context, the matrix $H$ is a Hamiltonian.
The number of bases in the Krylov subspace ($\nu$)
is chosen to be much smaller than that of the original Hamiltonian matrix $H$.
In a practical method for large-scale calculations,
we consider the Hamiltonian matrix 
{\it only within the above subspace},
which means the drastic reduction of the matrix size.  
The Krylov subspace gives the mathematical foundation 
of many numerical algorithms 
such as the standard conjugate gradient method\cite{KRYLOV} 
and the recursion method\cite{HAYDOCK-REV,PETTIFOR89}.

Recently, we developed 
a practical Krylov subspace method 
for the calculation of the density matrix
and applied it to molecular dynamics simulations \cite{TAKAYAMA}.
The number of bases in the subspace was chosen, typically, 
as $\nu = 30$.
We compared 
the resultant density matrix 
with that by  the Wannier state method.
Now we are planning 
other molecular dynamics simulations, 
especially, to metals.

\section{Hybrid scheme and parallel computations}

As another fundamental methodology for large-scale calculations,
we developed the hybrid scheme within quantum mechanics \cite{Hoshi02b,Hoshi03a}.
The basic idea is the followings;
The one-body density matrix is decomposed into two partial 
matrices or \lq subsystems' that are constructed 
from several occupied wave functions.
This decomposition corresponds to dividing the occupied Hilbert space.
The {\it different} partial density matrices are solved 
by {\it different} solver methods.
Each subsystem is obtained with a well-defined mapped Hamiltonian 
and a well-defined electron number\cite{Hoshi03a}.
Test calculations are
done by the combinations between 
(a) the  diagonalization method and perturbative Wannier state methods,
(b) the variational and perturbative Wannier state methods \cite{Hoshi02b}
(c) the Krylov subspace method and 
the perturbative Wannier state method \cite{Hoshi03a}.
Since the present hybrid scheme is a technique 
in calculating the density matrix $\hat{\rho}$,
any physical quantity is 
quantum mechanically well defined 
with Eq.~(\ref{TRACE-EQ}).

Parallel computation is also important 
for large-scale calculations.
Test calculation of
the perturbative Wannier state method is 
carried out
with upto 10$^6$ atoms  \cite{GESHI}
using the Message Passing Interface technique\cite{MPI-URL}
and 
with upto 10$^7$ atoms \cite{Hoshi03a}
using the OpenMP technique \cite{OMP-URL}.
We are now developing 
the parallelization of other methods
\cite{TAKAYAMA,Hoshi03a}.

\section{Application and discussion}

As a practical nanoscale application, the molecular dynamics simulation 
is performed for fracture of nanocrystalline silicon \cite{Hoshi02b}.
A standard workstation is used for the simulations
with upto 10$^5$ atoms.
We use the hybrid scheme 
between the variational Wannier state method 
and the perturbative Wannier state method.

In the continuum theory of fracture \cite{GRIFFITH,NOTE2},
a critical crack length 
is defined by a dimensional analysis
between the competitive energy terms of
the bulk strain (3D) energy 
and the surface formation (2D) energy.
Since the definition of the critical length is 
independent on the sample size or the lattice constant,
we can expect a crossover in fracture phenomena 
between nanoscale and macroscale samples.
The investigation of 
the above crossover is one purpose of the present simulation. 
Another purpose is 
the fracture behavior in atomistic pictures,  
on the points of how and why the fracture path 
is formed and propagates in the crystalline geometry \cite{NOTE-FRAC}.

Dynamical fracture processes 
are simulated under external loads in the [001] direction.
As the elementary process in fracture,
we observe a two-stage surface reconstruction process.
The process contains  
the drastic change of the Wannier states
from the bulk ({\it sp}$^3$) bonding state
to surface ones.
Figure \ref{FIG-ANIME} shows a result, 
in which the fracture propagates anisotropically on 
the (001) plane and reconstructed surfaces 
appear with asymmetric dimers \cite{Hoshi02b}. 
Step structures are formed in larger systems 
so as to reduce the anisotropic surface strain energy 
within a flat (001) surface.
Such a step formation is understood as
the beginning of a crossover 
between nanoscale and macroscale samples \cite{Hoshi02b}.
Further investigation should be done
for direct discussion of the crossover.

The present calculations are carried out 
using tight-binding Hamiltonian 
within $s$ and $p$ orbitals.
We should say that its  applicability is rather limited,
due to the simplicity of Hamiltonian.
Its parameter theory, however,
reproduces systematically 
several {\it ab initio} results 
among different elements or phases,
because the tight-binding formulation 
is universal within the scaled length and energy units
\cite{Hoshi00a,Hoshi01,Hoshi03a}.
An important future work is 
to construct simple and practical 
(tight-binding) Hamiltonians 
more systematically 
from the {\it ab initio} theory.
We will use the muffin-tin orbital formulation for the construction,
because it gives directly the tight-binding formulation
\cite{LMTO-HIGHLIGHT}.

Recently the concept \lq multiscale mechanics' is focused 
as the seamless theoretical connection 
of material  simulation methods
among the three principles of mechanics;
(I) quantum mechanics  (for electron systems), 
(II) classical mechanics, and 
(III) continuum mechanics.
The present work gives
a guiding principle and a typical example for the concept,
which is carried out  by simplifying the total energy functional.

\appendix
\section{Derivation of Conventional Wannier state \label{APPEND}}

Here we derive the conventional Wannier state
\cite{WANNIER-ORG,NOTE-AM}
as a specific  case of Eq.~(\ref{WANI-REV-LOC-ST-UT}).
In periodic systems,
eigen states are called Bloch states 
$\{ \psi_{\nu \bm{k}}^{\rm (Bloch)} \}$
with the suffices of 
the band $\nu$ and the k-point $\bm{k}$, 
the point in the Brillouin zone.
Within an isolated single band,
the Wannier states can be defined 
$W_{\nu \bm{l}}$ 
with the suffices of 
the band $\nu$ and the lattice vector $\bm{l}$;
\begin{eqnarray}
 W_{\nu \bm{l}} (\bm{r}) = \int \, d \bm{k} \, e^{-i \bm{k} \bm{r}} 
    \psi_{\nu \bm{k}}^{\rm (Bloch)} (\bm{r})
  \label{WANI-REV-ORIG-FORM},
\end{eqnarray}
where the integration is done within the Brillouin zone \cite{NOTE4}.
Equation (\ref{WANI-REV-LOC-ST-UT})
will be reduced to 
Eq.~(\ref{WANI-REV-ORIG-FORM}) ,
when 
the corresponding unitary matrix $U$ is chosen as
\begin{eqnarray}
 U_{ij} \Rightarrow U_{\nu \bm{l}, \nu' \bm{k}}
 \equiv \delta_{\nu \nu'} \, e^{-i \bm{k} \bm{r}}.
\end{eqnarray}
The conventional Wannier state
is given by 
the unitary transform 
only within an isolated single band $(\nu = \nu')$,
while the generalized Wannier states
are given by the unitary transform 
within different bands $(\nu \ne \nu')$.
It is also noteworthy that 
the concept of the generalized Wannier state,
unlike the conventional one,
can be applicable to non-periodic cases.

\newcommand{\noop}[1]{}


\begin{thebibliography}{10}
\bibitem{NOTE4}
In general,
Bloch states are not unique 
with respect to 
arbitrary phase freedoms 
($\phi^{\rm (eig)}_{\nu k}(\bm{r}) \rightarrow 
\phi^{\rm (eig)}_{\nu k}(\bm{r})
\exp (i \delta_{\nu k}) $).
This phase freedom, 
however, 
must be properly chosen
to construct localized states
in Eq.~(\ref{WANI-REV-ORIG-FORM}).
\end{thebibliography}
\end{document}